# Quantum Fourier-transform infrared spectroscopy in the fingerprint region

Y. Mukai,[1] R. Okamoto,[1,2] and S. Takeuchi[1,*]

[1] *Department of Electronic Science and Engineering, Kyoto University, Kyotodaigakukatsura, Nishikyo-ku, Kyoto 615-8510, Japan*
[2] *Japan Science and Technology Agency, PRESTO, Gobancho, Chiyoda-ku, Tokyo102-0076, Japan*
*takeuchi@kuee.kyoto-u.ac.jp

**Abstract**

Harnessing the quantum interference of photon-pair generation processes, infrared quantum absorption spectroscopy (IRQAS) can extract the infrared optical properties of a sample through visible or near-infrared photon detection without the need for an infrared optical source or detector, which has been an obstacle for higher sensitivity and spectrometer miniaturization. However, experimental demonstrations have been limited to wavelengths shorter than 5 μm or in the terahertz region, and have not been realized in the so-called fingerprint region of 1500–500 $cm^{-1}$ (6.6 to 20 μm), which is commonly used to identify chemical compounds or molecules. Here we report the experimental demonstration of quantum Fourier transform infrared (QFTIR) spectroscopy in the fingerprint region, by which both absorption and phase spectra (complex spectra) can be obtained from Fourier transformed quantum interferograms obtained with a single pixel visible-light detector. As demonstrations, we obtained the transmittance spectrum of a silicon wafer at around 10 μm (1000 $cm^{-1}$) and complex transmittance spectrum of a synthetic fluoropolymer sheet, polytetrafluoroethylene, in the wavelength range of 8 to 10.5 μm (1250 to 950 $cm^{-1}$), where absorption due to symmetric and asymmetric stretching modes of C-F bonds is clearly observed. The signal-to-noise ratio per unit spectral width and unit probe light intensity of our QFTIR spectroscopy method outperforms conventional FTIR spectroscopy by a factor of $10^2$. These results open the way for new forms of spectroscopic devices based on quantum technologies.

## I. INTRODUCTION

Spectroscopy is a versatile tool in a broad range of sciences from astronomy to life sciences. Spectroscopy in the far infrared (FIR) region, in particular in the so-called "fingerprint region" in the wavelength range of approximately 1500–500 $cm^{-1}$ (6.6 to 20 μm) [1], where complex patterns of absorptions are found that are unique to the chemical compounds or molecules, is of particular importance and is an essential tool for material science, chemistry, pharmaceutical sciences, and life sciences. As the most popular technique, Fourier transform infrared (FTIR) spectrometers have been widely used [2-4]. However, far-infrared light sources using a heating element and the lack of a highly sensitive detector with a wide dynamic range in the FIR region are critical bottlenecks for improving sensitivity and miniaturizing spectrometers.

Remarkable progress has been made in photonic quantum technologies [5], namely the applications of quantum entangled photons for quantum information [6-8], quantum communications [9-11], and quantum sensing [12-19]. Infrared quantum absorption spectroscopy (IRQAS), which harnesses the quantum interference of photon pair generation processes, enables spectroscopy in the infrared wavelength region using a visible light source to generate frequency entangled photon pairs in the infrared region and the visible and near-infrared (VNIR) region (0.4–1.1 μm, which can be detected by commonly used silicon photodetectors) and thus is free of the bottlenecks of the current IR spectrometers. Prompted by the pioneering demonstration of the IRQAS of $CO_2$ in the wavelength region around 4.2–4.5 μm [20], many important experiments [21-26] including the application of infrared microscopy [27,28] have been reported. However, these demonstrations were performed for wavelengths below 5 μm or in the terahertz region with wavelengths of 300 μm [22] and 600 μm [26]; IRQAS in the fingerprint region has not yet been demonstrated. A possible reason for this is that the nonlinear crystals (e.g., $LiNbO_3$, $LiTaO_3$, $BaB_2O_4$) used for entangled photon-pair generation are not transparent in the fingerprint region.

Here we report the experimental demonstration of quantum Fourier transform infrared (QFTIR) spectroscopy in the fingerprint region [23-25]. In this method, a spectrum in the IR region can be obtained by a Fourier transformation of an interferogram obtained with a single pixel silicon detector. Note that the interference in the entangled photon-pair generation process occurs in a so-called "low gain regime," where the average photon flux is below one photon pair per correlation time for the biphotons, which cannot be described by classical theory [29-32]. In addition to the ordinal transmittance spectra, QFTIR spectroscopy can provide the dispersion (refractive index spectra) of the sample [24]. To extend the operating wavelength of IRQAS to the fingerprint region, we newly developed a VNIR-FIR photon-pair source based on a spontaneous down-conversion process in an $AgGaS_2$ (AGS) crystal, which has a high transparency over a wide spectral range of 0.5 to 12 μm (20000 to 833 $cm^{-1}$). We successfully obtained the transmittance spectrum of a silicon wafer at around 10 μm (1000 $cm^{-1}$) and complex transmittance spectrum of polytetrafluoroethylene (PTFE) sheet in the wavelength range of 8 to 10.5 μm (1250 to 950 $cm^{-1}$), where absorption due to symmetric and asymmetric stretching modes of C-F bonds is clearly observed. The signal-to-noise ratio per unit spectral width and unit probe light intensity of our QFTIR spectroscopy method outperforms conventional FTIR spectroscopy by a factor of $10^2$. These results not only prove the applicability of IRQAS in the fingerprint region, but pave the way to new microscale devices and technologies for molecular/chemical contents recognition on the silicon-based integrated circuit platform [33,34] and also IR spectroscopy unencumbered by background thermal photons [35].

## II. Methods

In this section, we briefly review the theoretical basis for photon-pair generation via the SPDC process and the measurement procedures for QFTIR spectroscopy.

### A. Wavelength-tunable photon-pair generation via SPDC process

In an IRQAS system, the entangled photon-pair source is a key component because the photon-pair generation band and flux determine the operating wavelength range and measurement precision, respectively. In this work, we use a type-I ($e \rightarrow oo$) collinear SPDC process in a negative uniaxial nonlinear crystal for photon-pair generation, where a pump light with extraordinary polarization ($e$) is down-converted into signal and idler photons with ordinary polarizations ($o$). The energy conservation and phase matching conditions for this SPDC process can be represented as follows:

$$\hbar\omega_p = \hbar\omega_s + \hbar\omega_i, \quad (1)$$

$$n(k_p, \theta_p)k_p = n_o(k_s)k_s + n_o(k_i)k_i \quad (2)$$

,

where $\omega$ is the angular frequency, $\hbar$ is Dirac's constant, $k$ is the wavenumber in a vacuum, and the subscripts p, s, and i refer to the pump, signal, and idler photons. $n(k,\theta_p)$ is the refractive index for the extraordinary light with wavenumber $k$, whose value depends on the propagation angle of the light against the optic axis of the nonlinear crystal $\theta_p$, and is represented as

$$n(k,\theta_p) = \left(\frac{\sin^2\theta_p}{n_e(k)} + \frac{\cos^2\theta_p}{n_o(k)}\right)^{-1}, \qquad (3)$$

where $n_o(k)$ and $n_e(k)$ are the main refractive indices for ordinary and extraordinary light, respectively. Figures 1 depict the relation between the incidence angle $\theta_{in}$ and the propagation angle $\theta_p$ of the pump light. Here, the cut angle of the nonlinear crystal $\varphi$ is defined as the relative angle between the optic axis and the crystal surface normal. For normal incidence (Fig. 1 (a)), $\theta_p$ coincides with $\varphi$. As shown in Fig. 1(b), $\theta_p$ can be varied by rotating the nonlinear crystal, that is, changing $\theta_{in}$. Taking into account the refraction of the incident light, the relation between $\theta_{in}$ and $\theta_p$ can be obtained from Snell's law, $n(k_p,\theta_p)\sin(\varphi-\theta_p) = \sin\theta_{in}$. Equations (2) and (3) suggest that the SPDC wavelengths, determined by the phase-matching condition, can be tuned by rotating $\theta_p$, as demonstrated in Ref. 36 in the MIR region using an LiNbO$_3$ crystal.

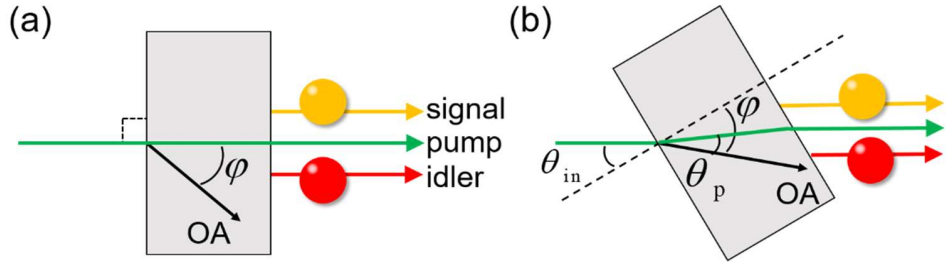

FIG. 1. Schematics of photon-pair generation via collinear SPDC process. (a) Normal incidence and (b) oblique incidence of the pump light on a nonlinear crystal. $\varphi$ is the cut angle, i.e., relative angle of the optic axis (OA) and the crystal surface normal. $\theta_{in}$ and $\theta_p$ are the incidece and propagation angles of the pump light, respectively.

To construct a VNIR–FIR photon-pair source, we chose an AgGaS$_2$ (AGS) crystal as the nonlinear crystal since it has high transparency over a wide spectral range of 0.5 to 12 µm and a large second-order nonlinear coefficient [37]. The refractive indices of AGS, $n_o(k)$ and $n_e(k)$, are modeled by the following Sellmeier dispersion equation [38]:

$$n_{o,e}^2(k) = A_{o,e} + \frac{B_{o,e}}{1 - C_{o,e}k^2} + \frac{D_{o,e}}{1 - E_{o,e}k^2}, \qquad (4)$$

where the Sellmeier coefficients for the ordinary and extraordinary light (labeled by the subscripts o and e) are $A_{o,e} = 3.3970, 3.5873$, $B_{o,e} = 2.3982, 1.9533$, $C_{o,e} = 0.09311, 0.11066$, $D_{o,e} = 2.1640, 2.3391$, and $E_{o,e} = 950.0, 1030.7$.

### B. Principle of QFTIR spectroscopy

QFTIR is a method of IRQAS where Fourier analysis is performed for the quantum interferogram and the complex transmittance of the sample is extracted. Here we follow the theoretical description for QFTIR spectroscopy considering a Michelson-type nonlinear quantum interferometer with a SPDC-based photon pair-source operated in the low-gain regime [24]. For the relative phase between related photons, which can be tuned through the optical path length difference $\Delta L$, the quantum interferogram $P_s(\Delta L)$, i.e., the signal photon count rate recorded as a function of $\Delta L$, can be given as

$$P_s(\Delta L) \propto 2 + \int_0^\infty dk_i |F_k'(k_i)|^2 (1-\gamma)\eta(k_i)\left(\left(\tau_k^*(k_i)\right)^2 e^{i\Theta} e^{-ik_i \Delta L} + \left(\tau_k(k_i)\right)^2 e^{-i\Theta} e^{ik_i \Delta L}\right), \quad (5)$$

where $|F'(k_i)|^2$ is a reduced expression for the two-photon field amplitude obtained from energy conservation and the monochromatic pump condition [24], $\tau_k(k_i)$ is the complex transmittance coefficient for a sample placed in the idler path, and $\eta(k_i)$ represents the effects of mode mismatch and the residual losses in the interferometer. Here, we represent the fixed phase terms independent of $\Delta L$ by $\Theta$. Basically, the visibility of the interferometric signal with spatial periodicity $k$ is determined by the transmittance of the sample at wavenumber $k$, $\tau_k(k)$. In general, the visibility of the quantum interference will be reduced due to wavefront distortion of the idler photons passing through the sample. In Eq. (5), we introduce this factor phenomenologically using the dephasing parameter $\gamma$, which we assume to be constant independent of the wavenumber. Note that for the quantum interferogram taken without the sample used for reference, $\gamma = 0$.

The Fourier transform of the quantum interferogram, $A_s(k) = (1/2\pi)\int d\Delta L P_s(\Delta L) e^{ik\Delta L}$, gives information about the spectral intensity of the idler photons and the transmittance of the sample:

$$A_s(k) \propto |F_k'(k)|^2 (1-\gamma)\eta(k)\left(\tau_k^*(k)\right)^2 e^{i\Theta}. \quad (6)$$

The transmittance spectrum (including the dephasing parameter) can be extracted by taking the ratio of $A_s(k)$ to a reference spectrum $A_s^0(k)$ taken without a sample $(\gamma(k) = 0$ and $\tau_k(k) = 1)$,

$$A_s(k)/A_s^0(k) = (1-\gamma)\left(\tau_k^*(k)\right)^2. \quad (7)$$

As discussed in Ref. 24, this analysis gives the complex transmittance spectrum, that is, both the magnitude of the transmittance and the phase spectrum of the sample. The dephasing in the sample, $\gamma$, can be evaluated when the transmittance at a certain wavenumber is known.

Also note that identification of the chemical compounds or molecules can be performed without the information about $\gamma$.

### III. Experimental setup

Figure 2 shows a schematic of the experimental setup for QFTIR spectroscopy with a Michelson-type nonlinear quantum interferometer. The pump source is a monochromatic continuous-wave (cw) laser diode with a center wavelength of 785 nm (output power: 100 mW, linewidth: < 100 MHz). The pump laser is focused by a lens L1 (f = 200 mm) into an AGS crystal (thickness: 0.5 mm, cut angle: 50°) where type-I SPDC generates VNIR–FIR photon pairs. The pump beam diameter is 1.2 mm before L1. The idler photons emitted from the AGS crystal are reflected by the dichroic mirror DM2 and spatially separated from the pump and signal photons. After collimation by an off-axis parabolic mirror OAPM (f = 150 mm), the idler photons pass through a germanium window G, where residual pump light and signal photons partially reflected by DM2 are filtered out. An end mirror $M_i$ in the idler path is placed on a high-precision translational stage (FS-1050UPX, Sigma Koki) to adjust the idler path length. The reflected idler photons are refocused onto the AGS crystal. A sample of interest is placed obetween G and $M_i$ in the idler path and the idler photons pass through the sample twice. The pump and signal photons are transmitted by DM2 and reflected by a concave mirror CM, and they are also refocused onto the AGS crystal. All the optical paths are aligned such that the spatial modes of the first SPDC photons perfectly match those of the second SPDC photons, to generate quantum interference between the photon-pair generation processes.

The signal photons from the quantum interferometer are transmitted by DM1 and collimated by a lens L2 (f = 200 mm). An iris I is placed after L2 and its aperture is set to 1.5 mm to spatially select only the center-most part of the signal beam where the SPDC emission angle is close to perfect collinear geometry. The signal photons are transmitted by a long-pass filter LPF and a short-pass filter SPF, while the residual pump light is filtered out. Then, the signal photons are coupled to a multi-mode fiber and sent to a detection system for quantum interferogram measurement. The flux of the signal photons is measured by a single photon counting silicon avalanche photodiode APD (SPCM-AQRH-14FC, Excelitas Tech., detection efficiency: 65 % at 700 nm, detectable wavelength range: VNIR (400 to 1060 nm)) and a photon counter (SR400, Stanford Research Systems). In this work, we define the quantum interferogram as the signal photon count rate $P_s$ recorded as a function of the optical path length difference between the signal and idler paths $\Delta L$. The signal photons are also sent to a spectrometer with a charge-coupled devise image sensor (SR500i + DU416A-LDC-DD, Andor, wavelength resolution: 0.1 nm) to measure the signal photon emission spectrum in the VNIR region.

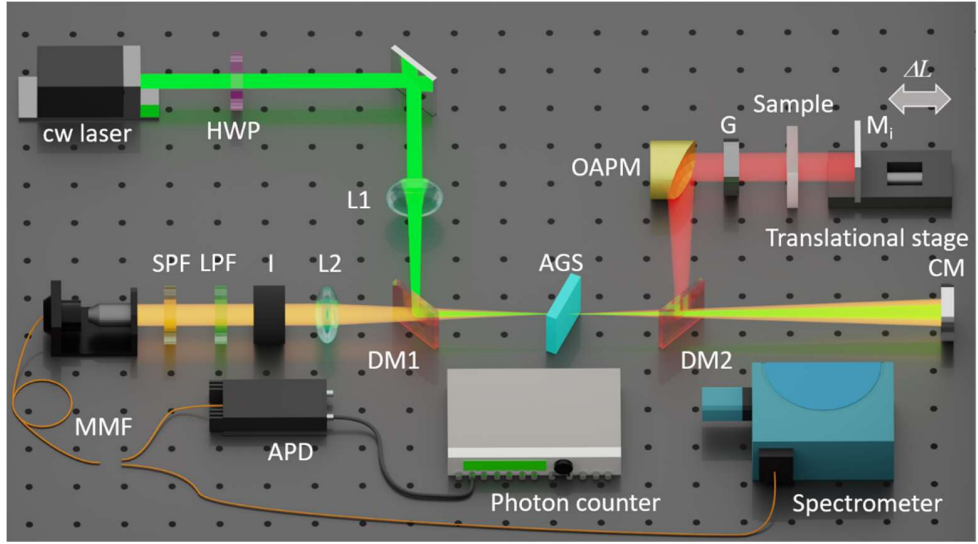

FIG. 2. Schematic of experimental setup. HWP: half wave plate, L1: focusing lens for the pump (*f* = 200 mm), L2: collimation lens for the signal (*f* = 200 mm), OAPM: off-axis parabolic mirror (*f* = 150 mm), DM1: long pass dichroic mirror (edge wavelength 800 nm), DM2: short pass dichroic mirror (edge wavelength 2000 nm), AGS: AgGaS$_2$ crystal, CM: concave end mirror, $M_i$: end mirror, LPF: long-pass filter (edge wavelength 800 nm), SPF: short-pass filter (edge wavelength 1000 nm), MMF: multi-mode fiber, APD: Si avalanche photodiode, G: germanium window, I: Iris. A continuous-wave (cw) laser with a wavelength of 785 nm (power 100 mW, linewidth < 100 MHz) is used as a pump.

Note that in this experiment, quantum interferometric measurements were performed in the low gain regime. The number of generated photon pairs from the crystal is estimated to be at most $5 \times 10^7$ pairs/s, considering the count rate of the signal photons $<1 \times 10^6$ counts/s and the effective total detection efficiency of 2 % (the quantum efficiency of the APD is more than 20 % over the signal photon generation band and the coupling efficiency of photons to the multimode fiber including optical loss in the system is at least 10 %). As a result, the estimated number of photon pairs per the biphotons' correlation time, about 100 fs, is $5 \times 10^{-6}$, which is much less than 1. Thus, we can conclude that our experiment shown in the next section is performed in the low gain regime.

## IV. Results and discussion

### A. Tunable VNIR-FIR photon-pair source

First, we confirmed that the idler photons are generated in the FIR region by measuring the emission spectra of the signal photons in the VNIR region. As the pump source is a monochromatic cw laser, the generation band of the idler photons can be uniquely determined from the wavelength of the signal photons, using energy conservation in the SPDC process (Eq. (1)). As explained in the previous section, the SPDC wavelengths can be selectively tuned by mechanical adjustment of the incidence angle of the pump beam to the nonlinear crystal $\theta_{in}$. Figure 3(a) shows signal photon spectra measured by the spectrometer with several different incidence angles $\theta_{in}$ by tilting the AGS crystal against the pump beam. For normal incidence, the wavelength of the signal photons is around 883 nm, suggesting idler photon generation around 7 μm (1430 cm$^{-1}$). The signal wavelength can be tuned from 845 to 900 nm by changing $\theta_{in}$ from −20° to 10°. This corresponds to the generation of idler photons in the FIR range of 6 to 11 μm (1670–910 cm$^{-1}$). The spectral intensities at $\theta_{in}$ of 5° and −10° are larger than the others but the spectral intensity at each angle is enough stable for a long period of time to perform spectroscopy measurement. In Fig. 3(b), the center wavelength for the signal photons is extracted by fitting a Gaussian curve to the spectral data in Fig. 3(a), and plotted as a function of the propagation angle of the pump light relative to the optic axis of the AGS crystal $\theta_p$, which can be calculated from $\theta_{in}$ using Snell's law as explained in Methods. The figure also shows the theoretical prediction of the SPDC wavelengths calculated from Eqs. (1)–(4). The experimental results show fair agreement with the theoretical calculation and suggest that our photon-pair source covers a wide FIR spectral range.

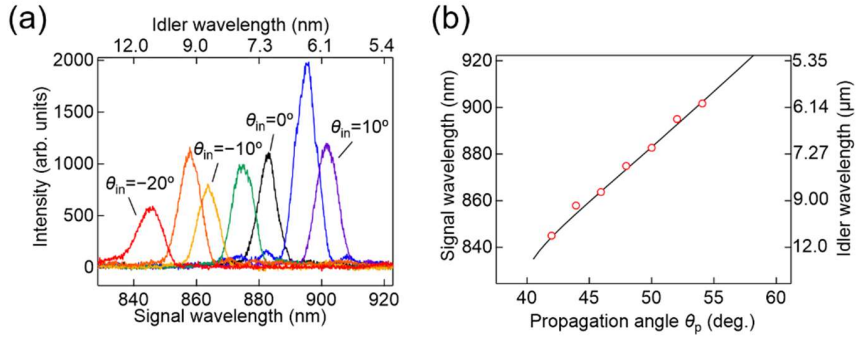

FIG. 3. (a) Signal emission spectra measured for several incident angles, plotted in 5° increments from 10° to −20°. The upper horizontal axis shows the corresponding idler wavelength. The incidence angle was changed by rotating the AGS crystal. (b) Center wavelengths of the signal emission spectra (red open circles) plotted as a function of the propagation angle of the pump light inside the AGS crystal against the optic axis. The right-side vertical axis shows the corresponding idler wavelength. The black line is the theoretical prediction of the phase-matched SPDC wavelength.

### B. Quantum interferogram in the fingerprint region

Utilizing this photon-pair source, we performed QFTIR spectroscopy in the fingerprint region. First, we measured a quantum interferogram without a sample. Here, the incidence angle of the pump light to the AGS crystal was adjusted to $\theta_{in} = -18°$ such that the signal and idler wavelengths were phase-matched at around 850 nm and 10 μm, respectively. The signal photon count rate measured by APD is shown in Fig. 4(a) as a function of $\Delta L$. We translated $M_i$ with step sizes of 500 nm (equivalent to a change in $\Delta L$ of 1000 nm) up to a total scanning length $W$ of 400 μm. The wavenumber resolution $\Delta k$ of the QFTIR measurement is determined by the inverse of the width of the transform window, that is, $\Delta k (= 1/W) = 25$ cm$^{-1}$. The origin of the optical path length difference is defined as the position where the interference fringe takes the maximum value. The photon count is integrated over 10 s for each step. The periodicity of the interferogram is estimated to be around 10 μm, as shown in the inset, being coincident with the expected generation wavelength of the idler photons. The visibility of the interferometric signal, $V = (P_{max} - P_{min})/(P_{max} + P_{min})$, was 11 %, where $P_{max}$ and $P_{min}$ are the maximum and minimum photon count rates observed in the quantum interferogram. A discrete Fourier transform (DFT) of the interferogram was performed along with a fast Fourier transform algorithm. The obtained Fourier amplitude spectrum is shown in Fig. 4(b). The spectral peak is located around 1000 cm$^{-1}$, corresponding to an idler wavelength of 10 μm, as expected, and the full-width of half maximum is about 100 cm$^{-1}$.

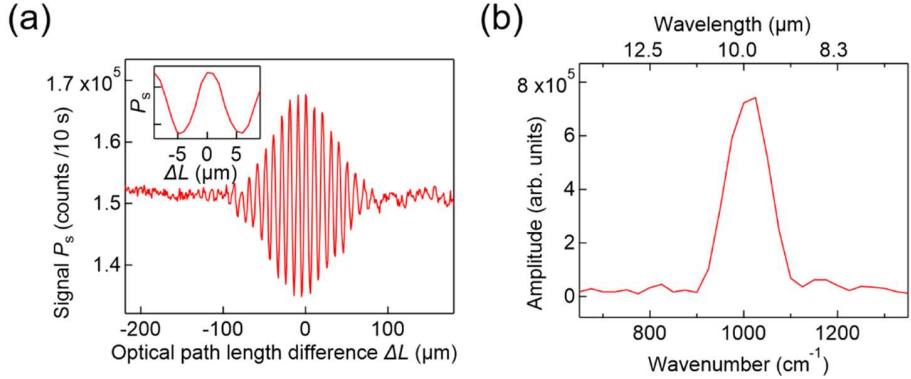

FIG. 4. (a) Quantum interferogram with signal and idler wavelengths of 850 nm and 10 μm, respectively. The inset shows an expanded plot around $\Delta L = 0$ μm. (b) Fourier amplitude spectrum of the interferogram.

### C. QFTIR spectroscopy for a silicon wafer in the fingerprint region

Next, as a demonstration of QFTIR spectroscopy in the FIR region, we measured the transmittance spectrum of a silicon wafer (thickness 1 mm) around 10 μm (1000 cm$^{-1}$). In this measurement, the total scanning length $W$ was set to 800 μm and the photon count was integrated over 500 ms for each step. The quantum interferograms were measured with and without the sample (Fig. 5(a)). Because of the high refractive index of silicon, a shift of about

5000 μm in the peak position was observed, along with a decrease in signal amplitude due to Fresnel reflection loss. Figure 5(b) shows the Fourier amplitude spectra of the inteferograms in Fig. 5(a). The wavenumber resolution $\Delta k$ is 12.5 cm$^{-1}$. Taking the ratio of these spectra, the transmittance spectrum of the silicon wafer is obtained as shown in Fig. 5(c). Here, the measurements were repeated 10 times and the average data for the transmittance spectra are plotted. In this spectral range, the obtained result shows an almost flat transmittance spectrum near 0.5. Since we used a polished optical-grade silicon wafer as the sample, we assumed the dephasing parameter $\gamma = 0$. For comparison, we also plot a transmittance spectrum measured by conventional FTIR spectroscopy (IRTracer-100, Shimadzu Corp.) with a wavenumber resolution of 4 cm$^{-1}$ in the same figure. The results of the QFTIR measurements are in good agreement with the data obtained from the conventional measurements.

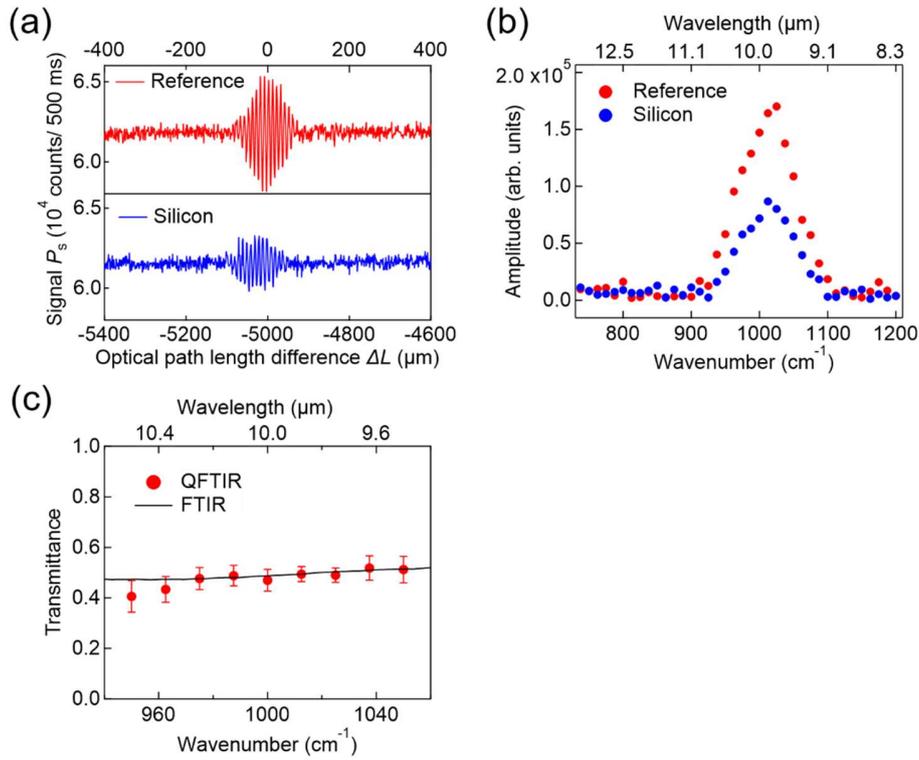

FIG. 5. (a) Reference quantum interferogram (red) and interferogram for silicon sample (blue). (b) Fourier amplitude spectra of signals in Fig. 5(a). (c) Transmittance spectrum of silicon wafer measured by QFTIR (red circles) and conventional FTIR (black line) spectroscopy. The error bars indicate standard deviations estimated from 10 measurements.

### D. QFTIR spectroscopy of a synthetic fluoropolymer sheet

Finally, we measured the complex transmittance spectrum of a polytetrafluoroethylene (PTFE) sheet, which has a steep spectral structure due to strong absorption lines in the FIR region. To cover a wide spectral range, QFTIR measurements were performed under several SPDC generation conditions by setting the incidence angle of the pump light $\theta_\text{in}$ to $-8$, $-13$, and $-18$ °. The total scanning length $W$ was set to 800 μm (wavenumber resolution $\Delta k$ of 12.5 cm$^{-1}$) and the photon count was integrated over 500 ms for each step. The reference and sample data for the quantum interferograms were taken at each crystal angle and the results are shown in

Figure 6(a). Figure 6(b) shows the reference spectra taken at each angle to clarify the spectral coverage of these measurements, showing that the Fourier spectra covered a wide spectral range from 950 to 1250 cm$^{-1}$ (corresponding to an FIR wavelength range of 8 to 10.5 μm).

The obtained QFTIR spectrum is shown in Fig. 6(c), where measurements were repeated 10 times for each SPDC generation condition and the average data for the transmittance spectra are plotted. Here we assumed a dephasing parameter of the sample $\gamma$ as 0.2, due to the surface roughness of the PTFE sheet. We also plot the transmittance spectrum measured with conventional FTIR spectroscopy for comparison. The transmittance spectrum obtained by QFTIR spectroscopy reproduces the spectral features of the PTFE sheet well, such as the sharp absorption dips around 1150 and 1200 cm$^{-1}$, which are assigned to symmetric and asymmetric stretching modes of the C-F bond [39]. Figure 6(d) shows the phase retardance spectrum, $\arg\left(A_s(k)/A_s^0(k)\right)$, which gives information about the dispersion of the sample. The phase spectrum is plotted in the rage of $\pm\pi$. The obtained result clearly exhibits a non-monotonic change in the phase spectrum, suggesting a large variation of the refractive index in this spectral range, which is expected from the strong optical absorption shown in Fig. 6(c). Note that Kramers-Kronig analysis is sometimes used to evaluate the dispersion of the material, however, generally we need to make strong assumptions regarding the optical response of the sample. In contrast, QFTIR spectroscopy can provide the dispersion of the sample directly from the experimental result without any assumptions.

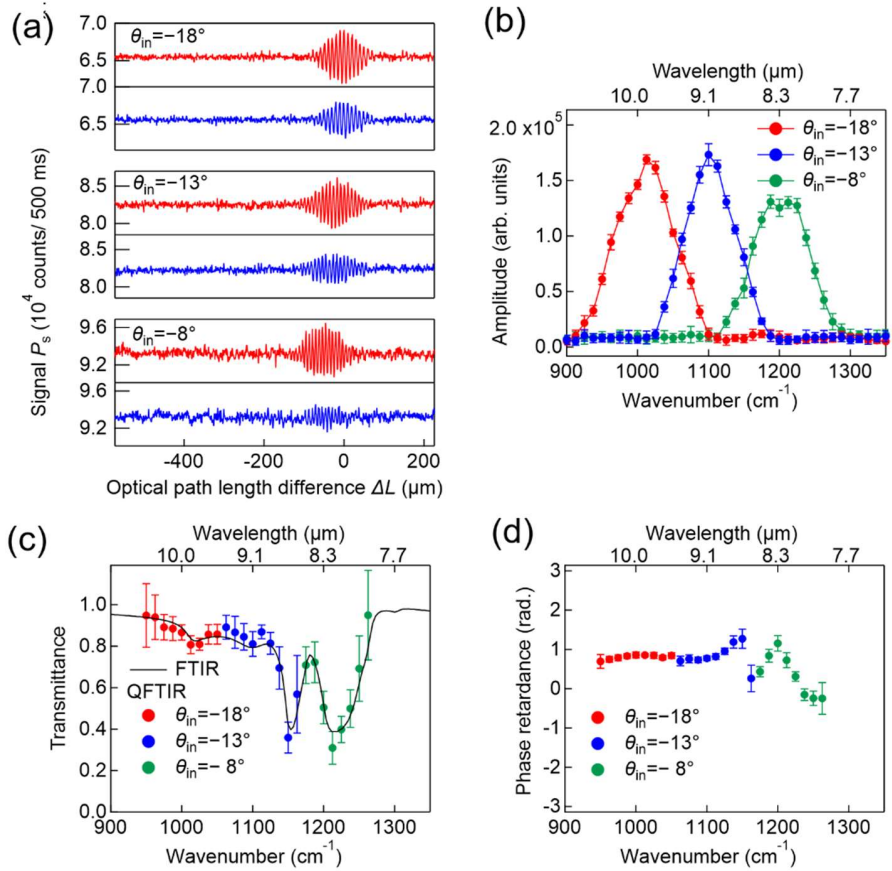

FIG. 6. (a) Reference quantum interferogram (red) and interferogram for PTFE sheet (blue) measured for different SPDC generation conditions. (b) Fourier amplitude spectra of reference signals in Fig. 6 (a). (c) Transmittance spectrum of PTFE sheet measured by QFTIR (red, blue, and green circles) and conventional FTIR (black line) spectroscopy. (d) Phase retardance spectrum. The error bars in (b), (c), and (d) indicate standard deviations estimated from 10 measurements.

### E. Signal-to-noise ratio

Here we discuss the signal-to-noise ratio (SNR) of the transmittance measurement by QFTIR and compare it to that in conventional FTIR measurements. Following the discussion in Ref. 25, we define the SNR as a ratio of the mean value of the transmittance measurement $T$ to the standard deviation $\delta T$,

$$SNR = T/\delta T . \qquad (8)$$

When the shot noise of the probe photon is the main source, $SNR$ is proportional to the square root of the number of probe photons $\sqrt{N}$ [40,41]. Hence, we introduce the following parameter $S$ as an index of the measurement precision per unit spectral width and unit probe light intensity,

$$S = SNR/\sqrt{N} . \qquad (9)$$

Here, $N$ is defined as the number of the photons passing through the sample whose wavenumber fall within a spectral range of $[k_0-(1/2)\Delta k, k_0+(1/2)\Delta k]$, where $k_0$ and $\Delta k$ is the detection wavenumber and spectral resolution of the transmittance measurement.

For QFTIR spectroscopy, $N$ is the number of idler photons in the FIR region which pass though the sample. $N$ is given as follows:

$$N = P \times t \times \Delta k , \qquad (10)$$

where $P$ is the idler photon flux passing through the sample per unit spectral width and $t$ is the measurement time required to get a single interferogram. We estimated the value of $SNR$ from the experimental results of the transmittance measurement for the PTFE sheet (Figs. 6) at $k_0 \approx 1000$ cm$^{-1}$, almost transparent region. $SNR_\text{QFTIR}$ and $N_\text{QFTIR}$ is approximately 23 and $4.2\times10^8$, respectively, and $S_\text{QFTIR}$ is calculated to be at least $1\times10^{-3}$ /photons$^{1/2}$.

Similarly, for conventional FTIR spectroscopy, $S_\text{FTIR}$ was estimated to be about $1\times10^{-5}$ /photos$^{1/2}$ by performing the transmittance measurements 10 times.

Comparing $S_\text{FTIR}$ and $S_\text{QFTIR}$, we conclude that the measurement precision of QFTIR spectroscopy is two orders of magnitude better than that for conventional FTIR spectroscopy given the same probe intensity per spectral width. While the conventional FTIR spectroscopy employs bright IR optical sources and sometimes causes optical damage to samples, QFTIR spectroscopy allows us to perform high-SNR spectroscopy with very-low photon flux. We think this feature of QFTIR is already highly advantageous for the non-invasive sensing of sensitive samples such as phototoxic molecules and delicate biological tissue.

The SNR of QFTIR measurements can be further improved by increasing the flux of the entangled photon pairs and the visibility of the quantum interferogram; the SNR of shot-noise limited Fourier spectroscopy is proportional to the visibility and the square root of the number of photons [40,41]. We estimate that the SNR of QFTIR spectroscopy becomes comparable to that of current FTIR spectroscopy with the same measurement time by increasing the photon flux by a factor of $6.7\times10^5$ when we do not consider the improvement in the visibility of the interferogram. The photon flux can be increased by more than one order of magnitude using a QPM device [42, 43] thanks to a long interaction length without spatial walk-off effect. Furthermore, waveguide structures that provide spatial confinement can also be used to improve the conversion efficiency; it has been reported that the photon-pair generation rate is enhanced by three orders of magnitude by using a waveguide QPM devices compared to a bulk QPM devices in the VNIR region [44]. An increased pump light intensity of 700 mW for a similar waveguide QPM device has been reported [45]. With these changes, the total photon flux can be enhanced by a factor of more than $7\times10^4$. In addition, by optimizing the mode matching, the visibility of the quantum interferogram is expected to be improved from the current value ($V$=11%) to $V > 80$ % as reported in [24], by which we can expect the improvement in the SNR by factor of 7.3 (equivalent to about 53-fold increase in the photon flux). With these improvements, it is possible that QFTIR spectroscopy outperforms current FTIR systems in the SNR.

## V. Conclusion

In conclusion, we have developed a QFTIR spectroscopy system for the FIR wavelength region, consisting of a Michelson-type nonlinear quantum interferometer with a VNIR–FIR photon-pair source based on the SPDC process in an AGS crystal. We measured the emission spectra of VNIR signal photons by varying the incidence angle of the pump laser relative to the AGS crystal, and confirmed that the signal wavelength can be tuned from 845 to 900 nm. This corresponds to the generation of idler photons over a wide FIR region from 6 to 11 μm (1670–910 cm$^{-1}$). Using this AGS-based photon-pair source, we successfully observed a quantum interferogram at an idler wavelength of 10 μm (1000 cm$^{-1}$). We measured the transmittance spectrum of a silicon wafer by QFTIR spectroscopy in this FIR region, and found that the obtained transmission spectrum was in good agreement with that obtained using conventional FTIR spectroscopy. The complex transmittance spectrum of a PTFE sheet was also measured over a wider spectral range of 8–10.5 μm (1250–950 cm$^{-1}$), changing the SPDC generation condition by angle tuning. The obtained results reproduced the steep absorption spectrum of this sample well and also extracted the phase retardance spectrum, which gives information about the dispersion. These results show that the operating wavelength of IRQAS can be extended to the fingerprint region, an infrared wavelength range important for spectroscopic applications, paving the way for the realization of broadband QAS systems as powerful tools for material identification and structural analysis. It has also been shown that the signal-to-noise ratio per unit spectral width and unit probe light intensity of our QFTIR method outperforms conventional FTIR spectroscopy by a factor of $10^2$. By employing a quasi-phase matching device and waveguide structure, we can expect that the flux of the photon-pair will be increased by more than four orders of magnitude. By using these photon-pair sources and constructing an optimal optical system that provides a high-visibility signal, it is possible that QFTIR spectroscopy outperform current FTIR systems in the SNR. In addition to the above methods, an on-chip photon-pair source, which is capable of generating broadband photon pairs in the IR region [33,34], is also a promising candidate for realizing high-precision IRQAS systems that can be integrated into silicon-based circuit platforms. We note that the nonlinear quantum interferometer operated at FIR wavelengths constructed in this work can be applied not only to QAS but also other quantum sensing applications such as hyperspectral imaging [27,28] and OCT in the infrared region [19].

## ACKNOWLEDGMENTS

The authors are grateful to H. Takashima and K. Shimazaki for their help in preparing the manuscript. This work was supported by the MEXT Quantum Leap Flagship Program (MEXT Q-LEAP) Grant Number JPMXS0118067634, JSPS KAKENHI Grant Number 21H04444, and JST-PRESTO (JPMJPR15P4).